\documentstyle[aaspp4]{article}

\begin{document}

\received{ 1999 August 4}
\revised{ 1999 September 10}
\accepted{1999 September ?}
\slugcomment{\shortstack[r]{
}
}

\lefthead{Kobayashi et al.}
\righthead{Near-infrared distant molecular cloud}

\title{Discovery of Young Stellar Objects\\
at the Edge of the Optical Disk of Our Galaxy}

\author{Naoto Kobayashi\altaffilmark{1}}
\affil{Subaru Telescope, National Astronomical Observatory of Japan\\
650 North A'ohoku Place, Hilo, Hawaii 96720}

\and

\author{Alan T. Tokunaga}
\affil{Institute for Astronomy, 2680 Woodlawn Dr., Honolulu, Hawaii 96822}

\altaffiltext{1}
{Visiting astronomer, Institute for Astronomy, 2680 Woodlawn Dr., Honolulu, Hawaii 96822}


\begin{abstract}        

  We report a discovery of young stellar objects associated with a
  molecular cloud at the edge of the optical disk of our Galaxy. This
  cloud is denoted as Cloud 2 in the list by Digel et al. and it is
  one of the most distant molecular clouds from the Galactic center
  known to date, with a probable distance of 15--19 kpc. We found
  seven red near-infrared sources associated with this cloud. Based on
  our near-infrared observations and far-infrared/radio data in the
  literature, we conclude that most sources are likely to be members
  of Cloud 2. The geometry of ionized gas, $IRAS$ sources,
  near-infrared sources, and molecular cloud suggests that MR-1, an
  isolated early B-type star near Cloud 2, has triggered the star
  formation activity in Cloud 2.

  Our results show that ongoing star formation is present in Cloud 2
  and that active star formation can occur in the farthest regions of
  the Galaxy, where the molecular gas density is extremely low,
  perturbation from the spiral arms is very small, and the metallicity
  is similar to that for irregular dwarf galaxies. Cloud 2 is an
  excellent laboratory in which to study the details of the star
  formation process in an environment that is similar to that in the
  early stage of the formation of the Galactic disk.

\end{abstract}  



\vspace{1cm}
\begin{center} {
{\large \it The Astrophysical Journal\\[2mm] 
received \underline{4 August 1999}, accepted \underline{16 September 1999}}
}
\end{center}

\newpage
\section{Introduction}
\label{sec : Introduction}

Digel, de Geus, \& Thaddeus (1994) found more than 10 molecular clouds
possibly beyond the optical disk of our Galaxy in the direction to the
Perseus arm (l $\sim$ 130$^\circ$). Their Galactic radius ($R_{\rm
  g}$) is estimated at more than 20 kpc and as much as 28 kpc (see
also, Digel et al. 1996; Heyer \& Terebey 1998). Because the
distribution of Population I and Population II stars in the Galaxy
have a sharp cutoff at around 18--20 kpc and 14 kpc, respectively
(Digel et al. 1994, and references therein), these distant molecular
clouds are potentially very interesting sites to investigate the
star-forming process away from the Galactic disk with little or no
perturbation from the spiral arms.

In such an outermost Galaxy region, the molecular gas surface density
is much smaller than in spiral arms (Heyer et al. 1998; Heyer \&
Terebey 1998; Digel et al. 1996) and the \ion{H}{1} surface density is
one fifth to one tenth of that in the spiral arms (e.g., Wouterloot et
al.  1990). Thus, the global star formation environment in the
outermost Galaxy region is quite different from that in the spiral
arms.  Also, metallicity is very low in such a region. The metal
abundance at $R_g$ = 20 kpc is estimated at 12 + log(O/H) $\sim$ 8.0,
assuming the standard abundance curve (e.g, Smartt \& Rolleston 1997).
This metallicity is comparable to that of dwarf irregular galaxies or
some damped Ly$\alpha$ systems of higher metallicity (see Ferguson,
Gallagher, \& Wyse 1998, and references therein). Therefore, studies
of star formation in the outermost Galaxy may reveal the details of
the star formation process in an environment similar to that thought
to exist during the early stage of the formation of the Galactic disk.

Of the 11 distant molecular clouds found by Digel et al. (1994, 1996),
Cloud 2 has the largest kinematic $R_{\rm g}$ of 28 kpc. If this
kinematic $R_{\rm g}$ is correct, Cloud 2 is located far beyond the
optical disk of our Galaxy and at the edge of the \ion{H}{1} gas disk
(around $\sim$30 kpc, e.g., Kulkarni, Blitz, \& Heiles 1982). Cloud 2
has a high CO luminosity (M$_{\rm CO}$ $\sim$ 3.7 $\times$ 10$^4$
$M_\odot$) that suggests star formation activity in this cloud (Digel
et al. 1994). Indeed, de Geus et al. (1993) found an extended
H$\alpha$ emission that has the same radial velocity as Cloud 2
($V_{\rm LSR}$ = $-$103 km s$^{-1}$). They concluded that the
H$\alpha$ traces an \ion{H}{2} region associated with Cloud 2, and
they proposed an early B-type star near Cloud 2 (``MR-1'': Muzzio \&
Rydgren 1974; see also Smartt et al.  1996) as the photoionizing
source. However, no star-forming activities like those in the nearby
star-forming molecular clouds have been reported thus far.

Smartt et al. (1996) obtained high spectral resolution optical spectra
of MR-1 and found $R_{\rm g}$ $\sim$ 15 to 19 kpc for MR-1 based on
the atmospheric parameters from their spectra and the photometry by
Muzzio \& Rydgren (1974)\footnote{ Smartt et al. (1996) derived an
  $R_{\rm g}$ of 15 kpc (heliocentric distance of 8.2 kpc) based on
  LTE model of the optical spectrum. They suggested that a non-LTE
  model can make it larger up to 19 kpc (heliocentric distance of 12
  kpc). Because the non-LTE model is more likely for stars like MR-1
  with high effective temperatures as described by Smartt et
  al. (1996), we assume $R_{\rm g}$ $=$ 19 kpc and a heliocentric
  distance of 12 kpc hereafter. The $R_{\rm g}$ of Earth is assumed to
  be 8.5 kpc.} . They suggest that MR-1 is probably physically related
to the \ion{H}{2} region (de Geus et al. 1993) because the radial
velocity of MR-1 ($-$90 $\pm$ 13 km s$^{-1}$) is in reasonable
agreement with the nebular velocity (--103 km s$^{-1}$).  If this is
the case, Cloud 2 is located close to the edge of the optical disk
($R_{\rm g}$ $\sim$ 20 kpc) rather than far beyond the optical
disk. However, it still remains as one of the most distant molecular
clouds/\ion{H}{2} regions known to date. The metal abundance of MR-1
is estimated at 12 $+$ log(O/H) $\sim$ 8.3 (Smartt \& Rollenston
1997), which is comparable to that for irregular dwarf galaxies (e.g.,
$\sim$8.4 for the Large Magellanic Cloud; Arnault et al.  1988).

Here we report a discovery of young stellar objects (YSOs) associated
with Cloud 2 made during our near-infrared studies. These sources
could shed light on the star formation processes in such a low-density
and low-metallicity environment as well as the distance to Cloud 2.
We have made comprehensive near-infrared observations of Cloud 2 that
include a wide-field survey, spectroscopy of detected infrared
sources, and deep imaging for the purpose of detecting low-mass
YSOs. The details of our study will be reported in subsequent papers.

\section{Observations and Results}
\label{sec : Observations and Results}

In October 1997, we made an initial near-infrared survey of Cloud 2
with University of Hawaii's QUIST (Quick Infrared Survey Telescope)
mounted at the UH 0.6 m telescope atop Mauna Kea. QUIST consists of
University of Hawaii's QUIRC (Quick Infrared Camera), a near-infrared
camera with 1024$\times$1024 HgCdTe HAWAII array, and 25.4 cm
Cassegrain telescope that provides a 25$\arcmin$ field of view with a
1.5$\arcsec$ pixel$^{-1}$ scale.\footnote{The 0.6-m telescope is not
  used. The QUIST telescope is attached to its equatorial mount.} The
observing was done remotely from the Institute for Astronomy in
Honolulu. The observations were partly affected by intermittent
cirrus.  Several standard stars from Elias et al. (1982) were observed
at several airmass positions for photometric calibration. We obtained
images of a field centered on Cloud 2 in three near-infrared bands,
$J$ (1.25\,$\mu$m), $H$ (1.65\,$\mu$m), and $K$ (2.2\,$\mu$m). The
total integration times for each band were 36 min, 36 min, and 45 min,
respectively.

We detected seven red sources associated with Cloud 2 with QUIST
(Fig. 1). The coordinates and near-infrared magnitudes of all
sources and MR-1 (Muzzio \& Rydgren 1974) are summarized in Table 1.
All of the near-infrared sources are associated with $IRAS$ sources in
Cloud 2: IRAS 02450+5816 for IRS 1; IRAS 02447+5811 for IRS 2, 3, 4,
5; and IRAS 02455+5808 for IRS 6\&7 (Fig. 2a and Table 2). {\it JHK}
photometry has been performed using IRAF APPHOT tasks.\footnote{ IRAF
  is distributed by the National Optical Astronomy Observatories,
  which are operated by the Association of Universities for Research
  in Astronomy, Inc., under cooperative agreement with the National
  Science Foundation.} An aperture of 18$\arcsec$ was employed. The
resultant $J-H$ vs. $H-K$ color-color diagram is shown in Figure 3.

We made follow-up $K$-band spectroscopy of IRS 1 and IRS 2 with the
near-infrared spectrograph CGS4 at UKIRT\footnote{The United Kingdom
  Infrared Telescope is operated by the Joint Astronomy Centre on
  behalf of the U.K. Particle Physics and Astronomy Research Council.}
in December 1997. IRS 1 and 2 are two bright sources near the northern
and southern clumps of the molecular cloud, respectively (see Fig. 2).
A 40 grooves mm$^{-1}$ grating that provides a spectral resolution of
$\lambda/\Delta\lambda$ $=$ 900 was used. Because seeing was
excellent, we used a narrow (0\farcs 6) slit with the tip-tilt
secondary. Observing conditions were photometric. To achieve
sufficient sampling, we took three exposures with one-third pixel
shift between exposures.  After basic reductions (e.g., sky
subtraction and flattening), one-dimensional spectra were extracted
with standard IRAF tasks. The standard star HR831 was used for the
correction of atmospheric extinction and flux calibration. The
Br$\gamma$ absorption line in the standard spectrum was removed by
interpolation before the extinction correction. The results are shown
in Fig. 4. When observing IRS1, the humidity was so low and stable
that we could clearly detect emission lines in spectral regions of
significant telluric water vapor absorption. The details of this
spectroscopy will be described in a separate paper with results from
our new CGS4 spectroscopy of additional Cloud 2 sources (Kobayashi \&
Tokunaga 1999a).

\section{Discussion}
\label{sec : Discussion}

\subsection{Near-infrared Sources}
\label{subsec : Near-infrared Sources}


Among all the detected sources in the observed field, the seven red
sources are distinctively red as shown in the true color pictures
(Fig. 1) and in the ($J-H$) vs. ($H-K$) color-color diagram (Fig.  3).
All of the red sources are associated with Cloud 2, and {\it no other
  bright red sources were found apart from Cloud 2 in the surveyed
  field} (Fig. 1). All sources except IRS 2 appeared to be point
sources with the QUIST spatial resolution ($\sim$2$\arcsec$).  IRS 2
is discussed in more detail in \S\ref{subsec : IRS 2}.

The Five College Radio Astronomy Observatory (FCRAO) $^{12}$CO survey
data (Heyer et al. 1998)\footnote{We obtained the FCRAO data
  electronically from the NASA Astronomical Data Center at
  http://adc.gsfc.nasa.gov/.} show no foreground clouds in the
direction of Cloud 2. However, a number of small foreground clouds are
around Cloud 2 in the surveyed field: a small cloud in the local arm
at 5$\arcmin$ southward, another small cloud in the local arm at
10$\arcmin$ westward, a small cloud associated with the Perseus arm at
10$\arcmin$ eastward, and a large cloud associated with the Perseus
arm at 20$\arcmin$ northward. In spite of the existence of many
foreground clouds in the surveyed field, we detected red sources only
in the small area centered at Cloud 2. This result strongly suggests
that all the red sources are physically associated with Cloud 2.


All seven sources show a large $H-K$ excess of more than 0.8. As shown
in Figure 3, the $H-K$ excesses of the sources except IRS 4 are not
due to interstellar extinction but are caused by intrinsically large
$H-K$ excess.
YSOs, highly obscured late-type stars (e.g, OH-IR stars,
protoplanetary nebulae (PPNs)), and active galactic nuclei (AGNs) are
known to show large intrinsic $H-K$ excess from dust emission (e.g.,
Lada \& Adams 1992 for YSOs; Garc\'\i a-Lario et al. 1997 for
late-type stars; and Hunt et al. 1997 for AGNs). Although it is
difficult to distinguish between these three classes of objects solely
from near-infrared colors, the red sources are most likely YSOs in
view of the association with the molecular cloud.


The two brightest red sources, IRS 6 and 7, which, among the red
sources, are most distant from Cloud 2 on the sky (Fig. 2a), might be
foreground stars in view of their brightness (a few magnitudes
brighter than other sources in Cloud 2; see Table 1) and relatively
large angular distance from Cloud 2 (7$\arcmin$--8$\arcmin$ from the
CO peaks; see Fig.  2a). Also, they are located at the edge of one of
the foreground molecular clouds in the Perseus arm. These two sources
are associated with the bright $IRAS$ point source, IRAS 02455+5808,
but not resolved within the $IRAS$ beam (1 $\sigma$ ellipse of 37$''$
$\times$ 10$''$ with PA $=$ 59$^\circ$). The $IRAS$ color is typical
for various kinds of objects (e.g., galaxies, YSOs, planetary nebulae)
and does not reveal the nature of IRS 6 and 7 clearly. The pointlike
appearance and extremely red near-infrared color ($H-K$ $>$ 1.5)
suggest that they are at least Galactic stars. 
Further study is necessary to clarify the nature of those sources.


MR-1 is located on a reddening vector from early-type stars. The
visual extinction of MR-1 is estimated at about $A_V =$ 3 to 4 mag
from this color-color diagram. This is consistent with the estimate
from $B$ and $V$ photometry of $A_V =$ 3.1 mag (Muzzio \& Rydgren 1974).

\subsection{IRS 1}
\label{subsec : IRS 1}


The $K$-band spectrum of IRS 1 (Fig. 4) shows three strong hydrogen
recombination lines: Pa$\alpha$ (1.875\,$\mu$m), Br$\delta$
(1.945\,$\mu$m), and Br$\gamma$ (2.166\,$\mu$m). Those lines show a
blueshift of about 100 to 200 km s$^{-1}$, suggesting IRS 1 is not an
extragalactic object.  Also, our $K$-band spectrum shows that IRS 1 is
unlikely to be an OH/IR star or a PPN because these objects do not
usually show hydrogen emission lines. OH/IR stars show strong
CO/H$_2$O absorption lines (Nagata 1999) and PPNs usually show
hydrogen {\it absorption} lines (Oudmaijer et al. 1995; Hrivnak, Kwok,
\& Geballe 1994). Although a few PPNs, possibly more evolved than most
PPNs, are known to show near-infrared hydrogen emission lines like
planetary nebulae (e.g., Aspin et al.  1993 for M 1-16; Thronson 1981
for AFGL 618), it is highly unlikely that such a rare source is
located near an $IRAS$ source in a molecular cloud (Fig. 2a). Instead,
it is highly plausible that the near-infrared emission lines are
signatures of an \ion{H}{2} region around YSOs.  For the reasons
above, we conclude that IRS 1 is a YSO physically associated with
Cloud 2.

Assuming the Galactic radius of 19 kpc for IRS 1 (heliocentric
distance = 12 kpc), the $K$-band absolute magnitude without any
correction for extinction is $M_K$ = $-$ 2.4 mag. This is comparable to
those for high to intermediate mass YSOs such as Herbig Ae/Be stars
(e.g., Hillenbrand et al. 1992). We estimate the spectral type of IRS
1 roughly at mid to late B from the K-band apparent magnitudes and
distances for the Herbig Ae/Be samples in Hillenbrand et al (1992).

\subsection{IRS 2}
\label{subsec : IRS 2}


IRS 2 is located at the southern peak of the CO molecular cloud as
well as at the center of the error ellipse of $IRAS$ 02450+5816 (Figs.
2a, 2b). IRS 2, 3, 4, and 5 form a cluster of red sources near the
southern CO peak (Figs. 2a, 2b); IRS 2 is the brightest.

The near-infrared color of IRS 2 shows that it is as highly
extinguished ($A_V$ $\sim$ 10 mag) as IRS 1. IRS 2 appeared to be
extended in the QUIST image with FWHM of $\sim$7$\arcsec$
(Fig. 1b). We recently obtained deep $JHK$ images of Cloud 2 with
higher spatial resolution and found that IRS 2 is a cluster of more
than 20 red pointlike sources (Kobayashi \& Tokunaga 1999b). This
morphology strongly suggests that IRS 2 is a star cluster in or behind
the molecular cloud. Further observations are necessary to clarify the
nature of IRS 2 as well as of IRS 3/4/5.

\subsection{Star Formation in Cloud 2}
\label{subsec : Star Formation in Cloud 2}


The ionized gas traced by H$\alpha$ emission extends from MR-1 toward
Cloud 2. The peaks of the H$\alpha$ emission are between the molecular
cloud peaks and MR-1 (de Geus et al. 1993; see also Fig. 2a). Since
IRS 1 is located at the center of the H$\alpha$ emission, it could
also be one of the major ionizing sources of this \ion{H}{2} region.
However, MR-1 is likely to dominate the ionization of the entire
\ion{H}{2} region because the number of ionizing photons from IRS 1 is
expected to be much lower than that from MR-1, assuming a spectral
type of mid- to late-B and B0--1, respectively, for IRS 1 and MR-1.


The $IRAS$ source associated with IRS 1 (02450+5816) is located
between the H$\alpha$ peak and northern CO peak of Cloud 2 (Figs. 2a,
2b). This pattern is typical for Galactic \ion{H}{2} regions (e.g.,
Gatley et al. 1979 for M17): young OB stars photoionize the surface of
an associated molecular cloud and make the warm dust region which is
traced by $IRAS$ 60/100$\mu$m flux.  IRAS 02450+5816 is not detected
at 12 or 25\,$\mu$m but only at 60 and 100\,$\mu$m. Its [60]-[100]
color temperature (about 30 K; assuming emissivity $\epsilon_\lambda$
$\sim$ $\lambda^{-2}$) is significantly lower than those for stars,
planetary nebulae, single YSOs or active galaxies (see, e.g., Walker
et al. 1989). Also, IRAS 02450+5816 is cataloged with a ``small-scale
structure flag,'' which denotes an association with a confirmed
extended source (IRAS Point Source Catalogue 1988). These
characteristics suggest that the $IRAS$ source is a warm extended
region adjacent to a molecular cloud rather than a single object with
compact far-infrared emission. Judging from the low dust temperature
($\sim$30 K), the warm region is not a prominent photodissociation
region (PDR) in a young star cluster (e.g., S140: Timmermann et
al. 1996) but a less energetic one in a dark cloud such as $\rho$ Oph
(Liseau et al.  1998). This is consistent with the suggestion by de
Geus et al. (1994) that Cloud 2 is more like a dark cloud (e.g.,
Taurus dark cloud) than a large star-forming complex with OB star
cluster (e.g., Orion molecular cloud complex).

The bolometric luminosity of IRAS 02450+5816 is estimated to be
$L_{\rm IR}$ $\sim$ 1000 $L_\odot$ from the $IRAS$ flux densities and
assuming a 12 kpc heliocentric distance (Emerson 1988; Tokunaga 1999).
Assuming a spectral type of B0V--B1V, the luminosity of MR-1 is
expected to be $L_{\rm IR}$ $\sim$ 10$^4$ $L_\odot$ if all the
emitting photons are entirely absorbed by the molecular cloud (see
Fig. 2 in MacLeod et al.  1998). If we assume that the northern peak
of Cloud 2 covers only 10\% of the sphere centered at MR-1 (the cone
of 60$^\circ$ apex angle), the observed $IRAS$ luminosity can be
explained by the ionization of MR-1. Although it is hard to estimate a
precise solid angle from the current data, it is likely that MR-1 is
the major ionizing source exciting the \ion{H}{2} region and the PDR.

The $IRAS$ source associated with the southern CO peak (IRAS
02447+5811) appears to have a small offset from the CO peak to MR-1 as
is the case for the northern peak (Fig. 2b). Since the geometry of an
ionizing source, ionizing gas, an $IRAS$ source, and a molecular cloud
is similar to that for the northern peak, IRAS 02447+5811 could also
be a PDR associated with Cloud 2.


The near-infrared sources IRS 1--5 are located between the H$\alpha$
peaks and the molecular cloud peaks near the $IRAS$ sources (Figs. 2a,
2b). This geometry suggests that the photoionization of MR-1 triggered
the formation of the near-infrared sources in Cloud 2.  Thus, the star
formation in Cloud 2 seems to be dominated by the single early B-type
star MR-1. It is also interesting to consider how a single B-star,
MR-1, was formed in the outermost Galaxy, but this is beyond the scope
of this paper.

\section{Conclusion}
\label{sec : Conclusion}

We have conducted a wide field near-infrared search for YSOs
associated with Cloud 2 as denoted by Digel et al. (1994). This cloud
is one of the most distant molecular clouds from the Galactic center
known thus far; the Galactic radius is estimated to be 15--19 kpc
(Smartt et al. 1996).
Although extended H$\alpha$ emission is associated with this cloud,
ongoing star-forming activity like that in the nearby star-forming
molecular clouds has not been previously reported.

We have discovered seven very red near-infrared sources in and around
Cloud 2 with wide-field imaging in the $J$ (1.25\,$\mu$m), $H$
(1.65\,$\mu$m), and $K$ (2.2\,$\mu$m) bands. Although foreground
clouds in Perseus and local spiral arms are around Cloud 2 on the sky,
we could not detect any red sources apart from Cloud 2 within the
total surveyed area of roughly 900 arcmin square.  Therefore, the
detected red sources are very likely to be members of Cloud 2.  Most
of the sources show large $H-K$ excess ($H-K$ $>$ 0.8), indicating
their YSO nature.  We have also obtained a $K$-band
(1.85--2.45\,$\mu$m) spectrum of two of the infrared sources, IRS 1
and IRS 2, that are near the two CO peaks in Cloud 2.  Strong hydrogen
emission lines (Br$\gamma$, Br$\delta$, and Pa$\alpha$) with a slight
blueshift were detected for IRS 1, while emission or absorption lines
were not detected for IRS 2 within the uncertainty. In view of the
cloud association and the emission-line spectrum, we conclude that IRS
1 is a YSO physically associated with Cloud 2.

IRS 1 is associated with an $IRAS$ point source with an extended
feature (IRAS 02450+5816) near the northern CO peak of Cloud 2. This
$IRAS$ source has a low color temperature ($\sim$30 K) and is located
between an H$\alpha$ peak and the CO peak, suggesting it is a
photodissociation region. IRS 2 is associated with IRAS 02447+5811 on
the southern CO peak of Cloud 2.  IRS 3, 4, and 5 are located around
this $IRAS$ source. The overall distribution of ionized gas, $IRAS$
sources, molecular cloud, and near-infrared sources suggests that
MR-1, an early B-type star near Cloud 2, has triggered the formation
of near-infrared sources in Cloud 2.


\acknowledgements 

We are grateful to Mike Nassir, Jim Deane, and Richard Wainscoat, and
to the staff of University of Hawaii 2.2 m telescope, for help during
our QUIST remote observing run. We thank the UKIRT support scientist
John Davis and the UKIRT staff for their kind help during our CGS4
observing run.  Special thanks goes to Tom Kerr of UKIRT for providing
special processing of the CGS4 data. Lastly, NK thanks Miwa Goto for
her useful comments on the first manuscript. NK was supported by a
JSPS overseas fellowship.



\begin{figure}
\figurenum{1a}
 \caption{ 
   True color image of the surveyed area from our QUIST data. Blue:
   $J$-band (1.25\,$\mu$m), Green: $H$-band (1.65\,$\mu$m), Red:
   $K$-band (2.2\,$\mu$m).  North is to the top, east is to the
   left. The image size is about 34$\arcmin$ (NS) $\times$ 40$\arcmin$
   (EW). Cloud 2 is located roughly at the center of this image, where
   the red sources are distributed. No red sources are seen except in
   a small area centered at Cloud 2.  At the four edges of this image,
   blue, red, and green colors are prominent because the images of
   each band do not cover the same area. Also, the random noise is
   higher near the edges because the number of co-added images are
   fewer than at the central region.}
\label{fig : figure1a}
\end{figure}

\begin{figure}
\figurenum{1b}
 \caption{ 
   Blow-up of the central region of Figure 1a. North is to the top,
   east is to the left. The image size is 23.0$\arcmin$ (NS)
   $\times$ 19.6$\arcmin$ (EW). This figure shows the same region on
   the sky as Figures 2a,2b.}
\label{fig : figure1b}
\end{figure}

\begin{figure}
\figurenum{2a} 
\caption{
  The seven red near-infrared sources overplotted on a grayscale
  $^{12}$CO map and contoured H$\alpha$ map from de Geus et al (1993).
  The asterisk ($*$) shows the location of MR-1. $IRAS$ sources are
  shown with a 1$\sigma$ error ellipse. IRS 1 is located at the center
  of the H$\alpha$ emission. IRS 2, 3, 4, 5 are closely packed near
  the southern CO peak where $IRAS$ 02447+5811 is situated, and IRS 2
  is at the center of the $IRAS$ error ellipse. IRS 6 and 7 are far
  from both the CO cloud and the H$\alpha$ emission.  The extended
  $IRAS$ source near IRS 1 (IRAS 02450+5816) is located between the
  H$\alpha$ peak and the northern CO peak of Cloud 2, suggesting it is
  a photodissociation region.}
\label{fig : figure2a}
\end{figure}

\begin{figure}
\figurenum{2b} 
\caption{
  The seven red sources on a grayscale $^{12}$CO map and $IRAS$
  60\,$\mu$m contour. The CO map was made from the data by Heyer et
  al. (1998) that was made available on the Internet. The CO flux was
  integrated over $v_{\rm LSR}$ = $-$98 to $-$108 km s$^{-1}$, which
  is the velocity range for Cloud 2. The gray-scale level ranges from
  2.4 (white) to 12 (black) K\,km\,s$^{-1}$. The $IRAS$ contour was
  made from the {\it High-Resolution IRAS Galaxy Atlas} (Cao et
  al. 1997): we obtained the $IRAS$ data from IPAC at
  http://www.ipac.caltech.edu/. Contours are 5, 10, 15, ... 40, 45,
  50, 80, 100 MJy sr$^{-1}$. }
\label{fig : figure2b}
\end{figure}

\begin{figure}
\figurenum{3} 
\caption{
  ($J-H$) vs. ($H-K$) color-color diagram for the Cloud 2 red sources
  and MR-1. Since IRS 4 was not detected in the $J$-band, only upper
  limit is shown for $J-H$. The main-sequence and giant star tracks
  and the reddening vector for $A_{\rm V} =$ 20 mag (arrow starting
  from the origin) are from Bessel \& Brett (1988).  The seven red
  sources are well on the right side of the reddening area for
  main-sequence and giant stars, showing their intrinsic $H-K$ excess.
  Note that the colors are in the ``QUIRC system''; color correction
  into more generic systems (e.g., CIT system) was not applied because
  the color conversion for the QUIRC system has not been
  established. The isochrones and the reddening arrow are in different
  filter systems (Bessel \& Brett 1988) and are shown for reference
  only.  }
\label{fig : figure3}
\end{figure}

\begin{figure}
\figurenum{4}
\caption{
  CGS-4 spectra of IRS 1 and IRS 2 in the $K$-band. The top solid bars
  show the wavelength ranges of significant atmospheric absorption.
  The spectrum of IRS 1 is offset by 1 mJy for clarity. Br$\gamma$
  (2.166\,$\mu$m), Br$\delta$ (1.945\,$\mu$m), and Pa$\alpha$
  (1.875\,$\mu$m) emission lines were clearly detected for IRS 1.
  H$_2$ emission (S(1) 1-0 at 2.122\,$\mu$m) or He I (2.058\,$\mu$m)
  lines were not detected. For IRS 2, we could not detect any emission
  lines within the uncertainty. The sharp emission feature at
  1.87$\mu$m for IRS 2 is an artifact from poor cancellation of
  atmospheric extinction.}
\label{fig : figure4}
\end{figure}


\begin{deluxetable}{cccccccccccccccc}
\scriptsize
\tablecolumns{16}
\tablewidth{0pc}
\tablecaption{List of Near-infrared Sources in Cloud 2}
\tablehead{
\colhead{Name} & \colhead{RA (2000)\tablenotemark{a}} & \colhead{Dec (2000)\tablenotemark{a}} & %
\colhead{$J$\tablenotemark{b}} & \colhead{$H$\tablenotemark{b}} & \colhead{$K$\tablenotemark{b}} & %
\colhead{$J-H$} & \colhead{$H-K$}  & %
\colhead{Associated $IRAS$ source}
}
\startdata

 IRS 1 &  2:48:56.52 &  58:29:20.2 & 15.39   (0.09) & 14.10   (0.08) & 12.99   (0.06) &  1.29   (0.12)  & 1.11   (0.10) & 02450+5816
\nl
 IRS 2 &  2:48:28.89 &  58:23:33.5 & 15.42   (0.09) & 14.25   (0.09) & 13.01   (0.06) &  1.17   (0.12)  & 1.25   (0.10) & 02447+5811
\nl
 IRS 3 &  2:48:27.15 &  58:23:57.0 & 16.92   (0.44) & 15.98   (0.55) & 14.81   (0.34) &  0.94   (0.71)  & 1.17   (0.65) & 02447+5811
\nl
 IRS 4 &  2:48:35.38 &  58:23:37.3 & \nodata        & 15.16   (0.16) & 14.09   (0.13) & \nodata         & 1.07   (0.21) & 02447+5811
\nl
 IRS 5 &  2:48:44.97 &  58:23:37.1 & 15.26   (0.07) & 14.08   (0.06) & 13.22   (0.06) &  1.18   (0.09)  & 0.86   (0.09)  & 02447+5811
\nl
 IRS 6 &  2:49:24.18 &  58:21:15.8 & 14.90   (0.05) & 12.97   (0.03) & 11.35   (0.01) &  1.93   (0.06)  & 1.61   (0.03)  & 02455+5808
\nl
 IRS 7 &  2:49:21.28 &  58:21:28.9 & 15.39   (0.07) & 13.34   (0.04) & 11.27   (0.01) &  2.05   (0.08)  & 2.07   (0.04)  & 02455+5808
\nl
 MR-1 &  2:49:22.35 &  58:26:44.6 & 11.49   (0.01) & 11.17   (0.01) & 10.95   (0.01) &  0.32   (0.01)  & 0.22   (0.01)  & \nodata
\nl
\enddata
\tablenotetext{a}{Accuracy is $\sim0\farcs2$, depending on the
  magnitude.}

\tablenotetext{b}{In the QUIRC photometric system. The statistical
  uncertainty from IRAF APPHOT task is in parentheses. The uncertainty
  shown does not include any systematic uncertainty from color
  transformation and imperfect weather conditions, which could be
  0.1--0.2 mag. }
\label{tbl : near-infrared sources}
\end{deluxetable}

\begin{deluxetable}{cccccc}
\scriptsize
\tablecolumns{16}
\tablewidth{0pc}
\tablecaption{List of $IRAS$ Sources in Cloud 2}
\tablehead{
&\multicolumn{4}{c}{Flux (mJy)} \\
\cline{2-5}
\colhead{Name} & \colhead{12\,$\mu$m} & \colhead{25\,$\mu$m} & %
\colhead{60\,$\mu$m} & \colhead{100\,$\mu$m} & %
\colhead{probable near-infrared counterpart}
}
\startdata

02447+5811 &  $<$0.27  & $<$0.25  & 0.72 & $<$4.47 & IRS 2,3,4,5
\nl
02450+5816 &  $<$0.31  & $<$0.25 & 1.81 & 8.85     & IRS 1
\nl
02455+5808 &  0.46 &  1.15 & 3.27 & 7.73             & IRS 6,7
\nl
\enddata
\label{tbl : iras sources}
\end{deluxetable}


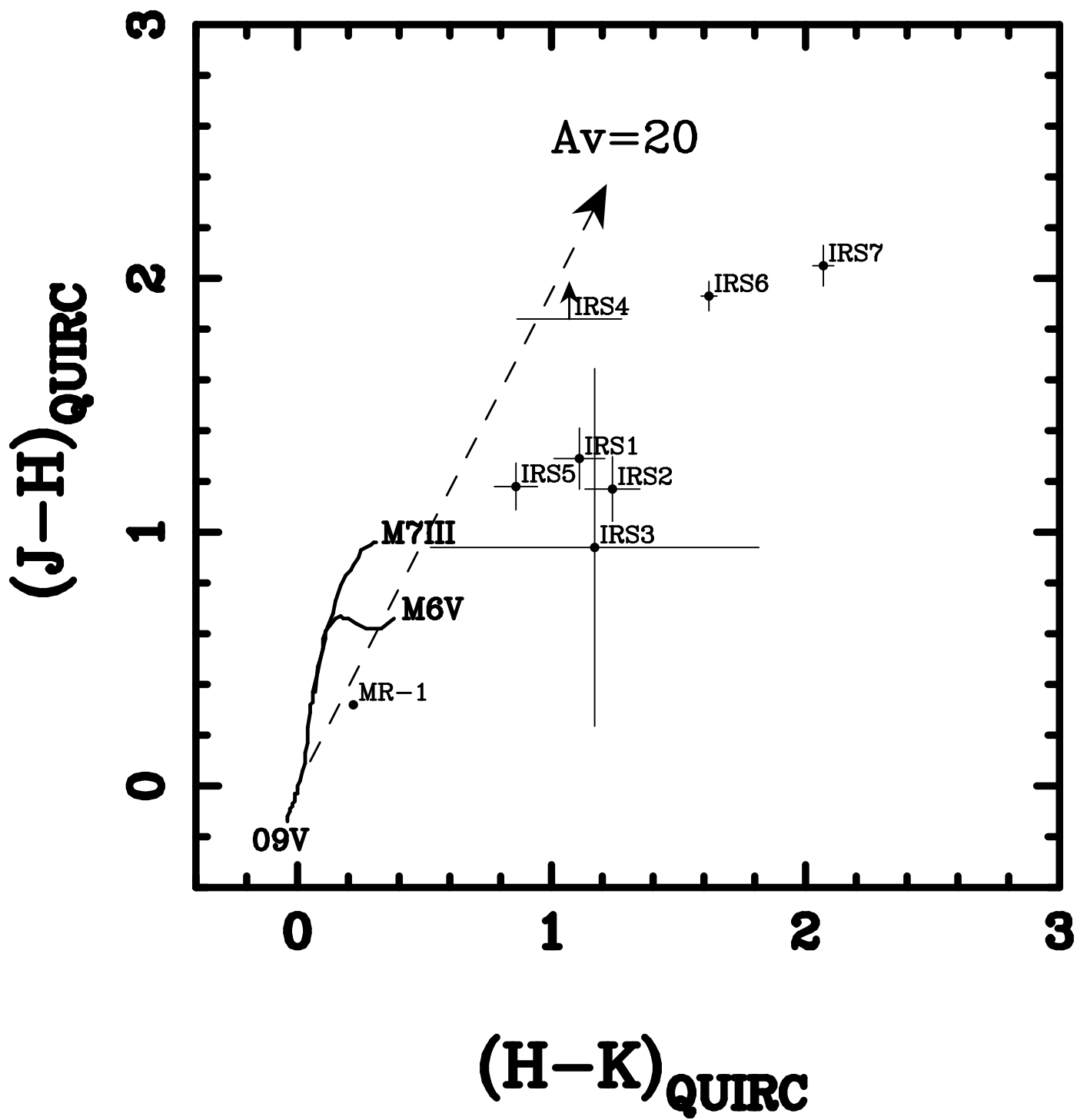
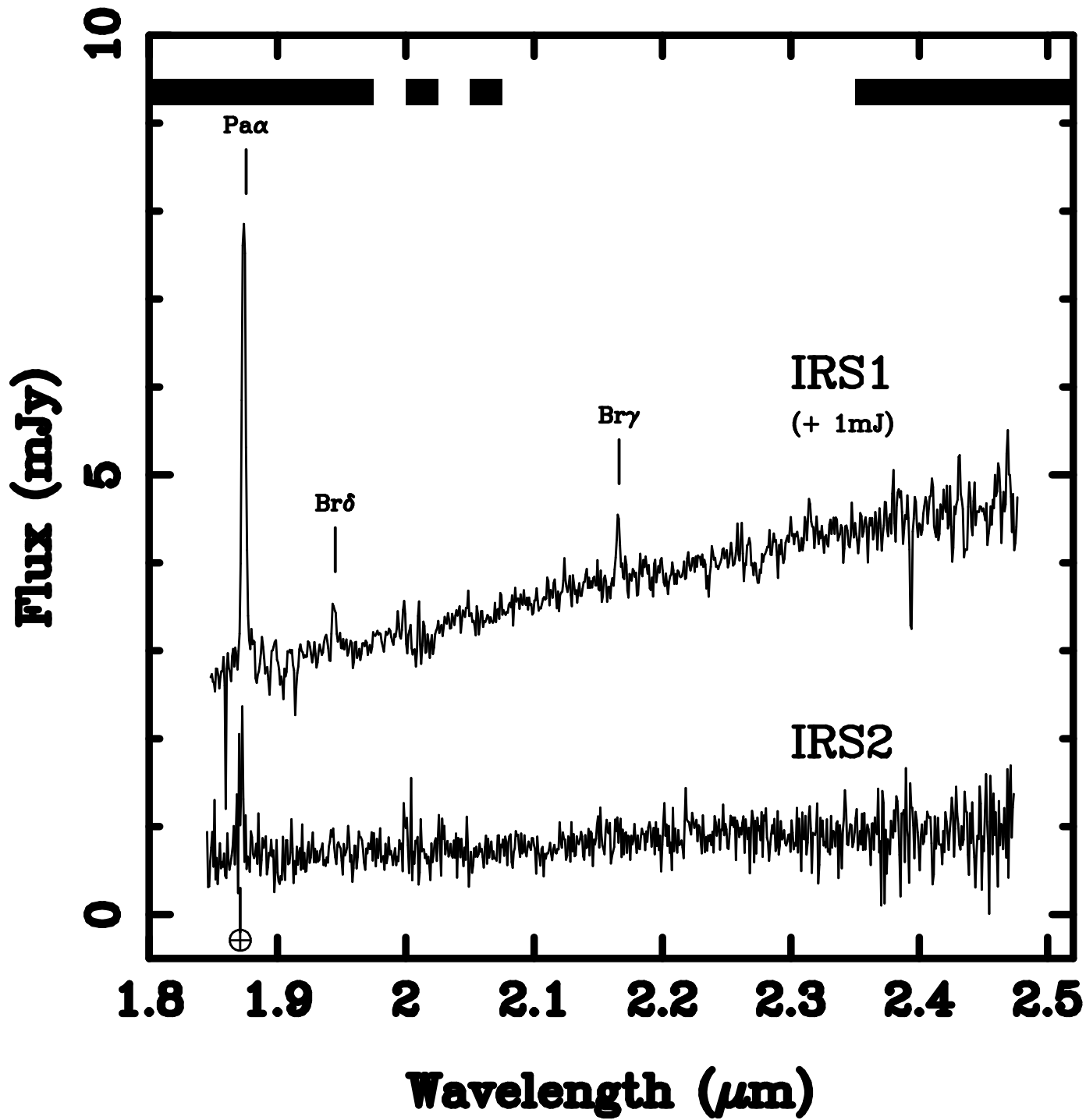
\end{document}